\documentclass[conference]{IEEEtran}
\usepackage{cite}
\usepackage{amsmath,amssymb,amsfonts}
\usepackage{algorithmic}
\usepackage{graphicx}
\usepackage{textcomp}
\usepackage{xcolor}
\def\BibTeX{{\rm B\kern-.05em{\sc i\kern-.025em b}\kern-.08em
    T\kern-.1667em\lower.7ex\hbox{E}\kern-.125emX}}
\begin{document}

\title{General Regression Neural Networks, Radial Basis Function Neural Networks, Support Vector Machines, and Feedforward Neural Networks}

\author{\IEEEauthorblockN{Alison Jenkins, Vinika Gupta, and Mary Lenoir}}

\maketitle

\begin{abstract}

The aim of this project is to develop a code to discover the optimal \begin{math}\sigma\end{math} value that maximum the F1 score and the optimal \begin{math}\sigma\end{math} value that maximizes the accuracy and to find out if they are the same. Four algorithms which can be used to solve this problem are: Genetic Regression Neural Networks (GRNNs), Radial Based Function (RBF) Neural Networks (RBFNNs), Support Vector Machines (SVMs) and Feedforward Neural Network (FFNNs).
\end{abstract}


\section{Introduction}

Based on the given data set, a second data set is generated. Any training instance with a desired output \begin{math}> 0.5\end{math} is relabeled as \begin{math}1.0\end{math}, and any training instance that has a desired output \begin{math}< 0.5\end{math} is relabeled as \begin{math}-1.0\end{math}. Using the two data sets, the GRNN, RBFNN, SVM and FFNN algorithms are implemented, and the statistics for each algorithm are recorded. 

\subsection{General Regression Neural Network}

For the GRNN algorithm, the Steady State Genetic Algorithm (SSGA) is used to evolve the \begin{math}\sigma\end{math} value. Apart from this, the training instances are also compared if the results of GRNN and EGRNN-I are similar\cite{b7}. 

\subsection{Radial Basis Function Neural Networks}

The RBFNN algorithm involves two methods to find out the optimal \begin{math}\sigma\end{math} value. One variation on the algorithm uses Kohonen Unsupervised Learning and Back-Propagation\cite{b6}. 


\subsubsection{\textit{Learning Vector Quantizer - I}}

The clustering algorithm constructs clusters of similar input vectors (patterns), where  similarity is usually measured in terms of Euclidean distance\cite{b1}. 
The training process of Learning Vector Quantizer-I (LVQ-I) is based on competition. 
During training, the cluster unit whose weight vector is the "closest" to the current input pattern is declared as the winner. The corresponding weight vector and that of the neighboring units are then adjusted to better resemble the input pattern\cite{b1}. 
It is not strictly necessary that LVQ-I uses a neighborhood function, thereby updating only the weights of the winning output unit\cite{b1}. 

\subsubsection{\textit{Learning Vector Quantizer - II}}

The Kohonen Learning Vector Quantizer - II (LVQ-II) uses information from a supervisor to implement a reward and punish scheme. LVQ-II assumes that the classifications of all input patterns are known. If the winning cluster unit correctly classifies the pattern, then the weights to that unit are rewarded by moving the weights to better match the input pattern. However, if the winning unit misclassified the input pattern, the weights are penalized by moving the weights away from the input vector\cite{b1}. 
Similarly to LVQ-I, a conscience factor can be incorporated to penalize frequent winners.


A RBFNN is a FFNN where hidden units do not implement an activation function, but represents a radial basis function. A RBFNN approximates a desired function by superposition of nonorthogonal, radially symmetric functions\cite{b1}. 
RBFNNs can improve accuracy and decrease training time complexity. The architecture of a RBFNN is similar to that of a FFNN, with the differences being that the hidden units implement a radial basis function, weights from the input units to the a hidden unit represent the center of the radial basis function, and some radial basis functions are characterized by a width, \begin{math}\sigma\end{math}. For such basis functions, the weight from the basis unit in the input layer to each hidden unit represents the width of the basis function. Note that the input unit has an input signal of \begin{math}+1\end{math}\cite{b1}. 
The output units of a RBFNN implement linear activation functions, so the output is simply a linear combination of basis functions. 
As with FFNNs, RBFNNs are universal approximators\cite{b1}. 

A limitation of K-Nearest Neighbors is that a large database of training examples must be kept in order to predictions. The LVQ-I algorithm allows learning with a much smaller subset of patterns that best represent the training data. 


\subsection{Support Vector Machines}

The SVM algorithm uses the derived data set that results from analyzing the given data set. This algorithm also has two parts: Linear SVMs where a linear kernel is used and Radial Basis SVMs where a Gaussian kernel is used. Both SVM algorithms are implemented using scikit-learn. scikit-learn is a Python package which is used to implement machine learning algorithms. In this case, \begin{math}90\end{math} percent of the data set is used as a training set, and the remaining \begin{math}10\end{math} percent is used as the test set\cite{b5}. 

\subsection{Feedforward Neural Network}

The FFNN algorithm uses scikit-learn to solve the problem using different numbers of hidden layers. The \begin{math}\sigma\end{math} value is determined using 1, 2 and 4 hidden layers. By convention, the hidden functions must be uniform. However, every hidden layer can use a different activation function, for example, a Gaussian activation function for neurons in the first layer, linear activation functions for neurons in the second layer, etc. The algorithm used in this paper is a sigmoidal activation function\cite{b5}. 

\section{Methodology}

The given data set is used to discover the \begin{math}\sigma\end{math} value for the GRNNs and RBFNNs and FFNN. The derived data set is used to discover the \begin{math}\sigma\end{math} value for the SVMs. Further analysis compares the performance of each of these algorithms and determines the best performing algorithm. 

%

\subsection{Radial Basis Function Neural Networks}

\begin{figure}
\begin{center}
\setlength{\unitlength}{0.012500in}
\includegraphics[width=90mm, scale=1.5]{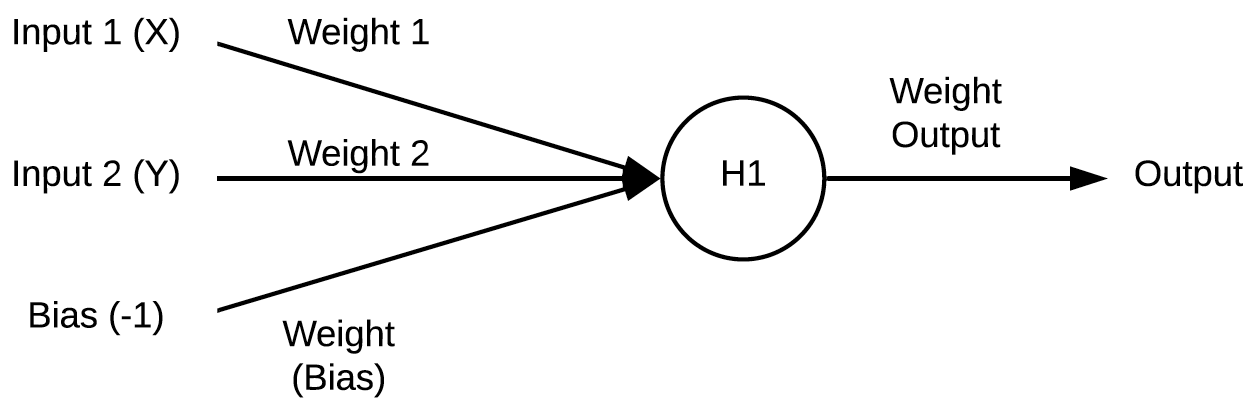}
\end{center}
\caption{Radial Basis Function Neural Network Topology}
\label{figure_RBFNN_topology}
\end{figure}

\begin{figure}
\begin{center}
\setlength{\unitlength}{0.012500in}
\includegraphics[width=90mm, scale=1.5]{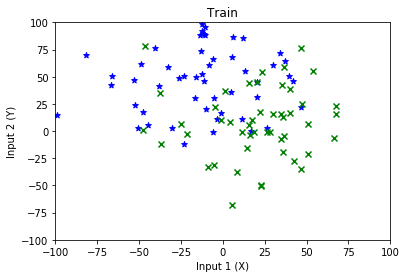}
\end{center}
\caption{Clustering Algorithm}
\label{figure_clustering_kohonen}
\end{figure}



The topology of the RBFNN is shown in Figure~\ref{figure_RBFNN_topology}. The clustering algorithm is shown in Figure~\ref{figure_clustering_kohonen}. First, the network weights, the learning rate, and the neighborhood radius are initialized. The Kohonen LVQ-I algorithm initializes the weights with random values, samples from a uniform distribution, or by taking the first input patterns as the initial weight vectors. Stopping conditions may be: a maximum number of epochs is reached, stop when weight adjustments are sufficiently small, or a small enough quantization error has been reached. While the stopping conditions are not true, for each pattern, the Euclidean distance is calculated. The output unit for which the distance is the smallest is found. Then, all the weights are updated for the neighborhood. After each pattern has had all of the weights updated, the learning rate is updated. Then, the neighborhood radius is reduced at specified learning iterations\cite{b5}. 


Each hidden unit implements a radial basis function, or kernel function, which is a strictly positive, radially symmetric function. A RBF has a unique maximum at its center, \begin{math}\mu\end{math}, and the function usually drops off to zero rapidly further away from the center. The output of a hidden unit indicates the closeness of the input vector, \begin{math}z_p\end{math}, to the center of the basis function. Some RBFs are characterized by a width, \begin{math}\sigma_j\end{math}, which specifies the width of the receptive field of the RBF in the input space for the hidden unit \begin{math}j\end{math}\cite{b5}. 

%

The Gaussian kernel is an example of radial basis function kernel, and is shown in Equation (\ref{eq_gaussiankernel_e1}). 

\begin{equation}
k(x,y) = \exp{({-\frac{||x-y||^2}{2\sigma^2}})}
\label{eq_gaussiankernel_e1}
\end{equation}


Algorithms for training RBFNNs can vary. Methods used to train RBFNNs differ in the number of parameters that are learned. The fixed centers algorithm adapts only the weights between the hidden and output layers. Adaptive centers training algorithms adapt weights, centers, and deviations. Gradient descent can be used to adjust weights, centers, and widths. Centers can be initialized in an unsupervised training step prior to training the weights between hidden units (radial basis) and output units\cite{b1}. 




Weight initialization can be based on gradient-based optimization methods. Gradient-based optimization methods, such as gradient descent, are very sensitive to the initial weight vectors. If the initial position is near the local minimum, then convergence will be quick. However, if the initial weight vector is on a flat area on the error surface, then convergence is slow. Also, large initial weight values may prematurely saturate units due to extreme output values with associated zero derivatives. In the case of optimization algorithms, such as Particle Swarm Optimization (PSO) and Genetic Algorithms (GAs), initialization should be uniformly distributed over the entire search space to ensure that all parts of the search space are covered\cite{b1}. 


Momentum can be based on stochastic learning. Stochastic learning, where weights are adjusted after each pattern presentation, has the disadvantage of fluctuating changes in the sign of the error derivatives. The network spends a lot of time going back and forth, unlearning what the previous steps have learned. Batch learning is a solution to this problem, since weight changes are accumulated and applied only after all patterns in the training set have been presented. Another solution is to keep with stochastic learning, and to add a momentum term\cite{b1}. 
The idea of the momentum term is to average the weight changes, thereby ensuring that the search path is in the average downhill direction. The momentum term is then simply the previous weight change weighted by a scalar value \begin{math}\alpha\end{math}\cite{b1}. 


%
%
%

\section{Experiment}

%
%

\subsection{Radial Basis Function Neural Networks}



Gradient Descent Training of RBFNNs begins with the selection of the number of centers. For each center, the center location, mean, and \begin{math}\sigma_j\end{math} values are chosen. Then, each weight is initialized. Next, a loop that runs until a stopping condition is reached computes the output, weight adjustment step size, and adjusts the weights. Lastly, the center step size is also computed, and the centers are adjusted. The \begin{math}\sigma_j\end{math} width step size is computed, and then the widths are adjusted\cite{b1}. 

Before the LVQ-I training phase, the RBFNN is initialized. The centers are initialized by setting all the \begin{math}\mu_{ji}\end{math} weights to the average value of all inputs in the training set. The weights are initialized by setting all \begin{math}\sigma_j\end{math} to the standard deviation of all input values of the training set. The hidden-to-output weights, \begin{math}w_{ki}\end{math}, are initialized to small random values. At the end of each LVQ-I iteration, the basis function widths are recalculated. For each hidden unit, the average of the Euclidean distance between \begin{math}\mu_{ji}\end{math} is computed, and the input patterns for which the hidden unit is selected is the winner. The width, \begin{math}\sigma_j\end{math}, is set to the average\cite{b1}. 


The training set provides data for the neural network to be trained on, and the bias value represents the threshold values of neurons in the next layer\cite{b1}. 
To simplify learning equations, the input vector is augmented to include and additional input unit, or the bias unit. The weight of the bias unit serves as the value of the threshold\cite{b1}. 
The strength of the output signal is influenced by the threshold value, or bias\cite{b1}. 

%
%
%

\section{Results}


The results of the statistical analysis of a comparison of all the algorithms is shown in Table~\ref{table_comparison_results}. 

\begin{table}[htbp]
\caption{Comparison Results}
\begin{center}
\begin{tabular}{|c|c|c|c|c|}
\hline
\textbf{}&\multicolumn{4}{|c|}{\textbf{}} \\
\cline{2-5} 
\textbf{Algorithm} & \textbf{\textit{Accuracy}} & \textbf{\textit{Precision}} & \textbf{\textit{Recall}} & \textbf{\textit{F1}} \\
\hline
GRNN & 0.4315 &  0.3693 &  0.2457 &  0.2571 \\
EGRNN & 0.4328 &  0.4163 &  0.3682 &  0.3369 \\
RBFNN$^{\mathrm{1}}$& 0.2367 &  0.3454 &  0.2335 &  0.2342 \\
RBFNN$^{\mathrm{2}}$& 0.2735 &  0.4245 &  0.3253 &  0.3726 \\
SVM$^{\mathrm{3}}$ & 0.4342 &  0.4137 &  0.4577 &  0.4643 \\
SVM$^{\mathrm{4}}$ & 0.3172 &  0.2845 &  0.3435 &  0.2281 \\
FFNN & 0.1293 &  0.1346 &  0.3946 &  0.2986 \\
\hline
\multicolumn{5}{l}{$^{\mathrm{1}}$Without Kohonen Unsupervised Learning and Back-Propagation} \\
\multicolumn{5}{l}{$^{\mathrm{2}}$With Kohonen Unsupervised Learning and Back-Propagation} \\
\multicolumn{5}{l}{$^{\mathrm{3}}$Linear SVM} \\
\multicolumn{5}{l}{$^{\mathrm{4}}$Radial Basis SVM} \\
\end{tabular}
\label{table_comparison_results}
\end{center}
\end{table}

%
%

\subsection{Radial Basis Function Neural Networks}

%

The results of the statistical analysis are shown in Table~\ref{table_rbfnn_results}. 

\begin{table}[htbp]
\caption{Comparison Results}
\begin{center}
\begin{tabular}{|c|c|c|c|c|}
\hline
\textbf{}&\multicolumn{4}{|c|}{\textbf{}} \\
\cline{2-5} 
\textbf{Algorithm} & \textbf{\textit{Accuracy}} & \textbf{\textit{Precision}} & \textbf{\textit{Recall}} & \textbf{\textit{F1}} \\
\hline
RBFNN$^{\mathrm{1}}$& 0.2367 &  0.3454 &  0.2335 &  0.2342 \\
RBFNN$^{\mathrm{2}}$& 0.2735 &  0.4245 &  0.3253 &  0.3726 \\
\hline
\multicolumn{5}{l}{$^{\mathrm{1}}$Without Kohonen Unsupervised Learning and Back-Propagation} \\
\multicolumn{5}{l}{$^{\mathrm{2}}$With Kohonen Unsupervised Learning and Back-Propagation} \\
\end{tabular}
\label{table_rbfnn_results}
\end{center}
\end{table}




A problem with LVQ networks is that one cluster unit may dominate as the winning cluster unit, thus putting most patterns in one cluster. To prevent one output unit from dominating, a "conscience" factor that penalizes an output for winning too many times may be incorporated in a function to determine the winning output unit\cite{b1}. 



The accuracy of a RBFNN is influenced by: the number of basis functions used, the location of the basis functions, and the width of the receptive field, \begin{math}\sigma_j\end{math}. A larger \begin{math}\sigma_j\end{math} represents more of the input space by that basis function. The larger the number of basis functions that are used, the better the approximation of the target function. However, the cost is an increase in computational complexity. The location of the basis functions are defined by the center vector, \begin{math}\eta_j\end{math}, for each basis function. Basis functions should be evenly distributed to cover the entire input space\cite{b1}. 

Training of a RBFNN should consider methods to find the best values for the parameters that affect the accuracy of the RBFNN. 
For the RBFNN without Kohonen Unsupervised Learning and Back-Propagation, the accuracy is 0.2367, while the precision is 0.3454. The recall is 0.2335, while the F1 value is 0.2342. 
For the RBFNN with Kohonen Unsupervised Learning and Back-Propagation, the accuracy is 0.2735, while the precision is 0.4245. The recall is 0.3253, while the F1 value is 0.3726. 

%
%
%

\section{Conclusions}

The aim of this project is to develop a code to discover the optimal \begin{math}\sigma\end{math} value that maximum the F1 score and the optimal \begin{math}\sigma\end{math} value that maximizes the accuracy and to find out if they are the same. Four algorithms which can be used to solve this problem are: GRNNs, RBFNNs, SVMs, and FFNNs. Based on the given data set, a second data set is generated. Any training instance with a desired output \begin{math}> 0.5\end{math} is relabeled as \begin{math}1.0\end{math}, and any training instance that has a desired output \begin{math}< 0.5\end{math} is relabeled as \begin{math}-1.0\end{math}. Using the two data sets, the GRNN, RBFNN, SVM and FFNN algorithms are implemented, and the statistics for each algorithm are analyzed and compared.

%
%

The RBFNN with Kohonen Unsupervised Learning and Back-Propagation performs better than the RBFNN without Kohonen Unsupervised Learning and Back-Propagation. The statistical comparison shows the RBFNN with Kohonen Unsupervised Learning and Back-Propagation results in a more accurate solution than that of the RBFNN without Kohonen Unsupervised Learning and Back-Propagation.




%
%
%

\section{Breakdown of the Work}

Vinika Gupta - SVM, Analysis, and {\LaTeX} Report (Abstract/Introduction/Methodology). 

Alison Jenkins - RBFNNs, GRNN, Analysis, and {\LaTeX} Report (Experiment, Results/Editing/Correct Formatting). 

Mary Lenoir - FFNN, Analysis, and {\LaTeX} Report (Final Edit, Format, References).

\vspace{12pt}
\color{red}

\end{document}